\documentclass[a4paper,11pt]{article}
\usepackage{pos}
\usepackage{pifont, bm}
\usepackage{tikz, pict2e}
\usepackage{centernot}
\usepackage{tkz-euclide}
\usepackage{float}
\usepackage{enumitem}

\usetikzlibrary{calc,backgrounds,intersections}
\usetikzlibrary{arrows,shapes,positioning,shadows,trees,shadows.blur,shadings}
\usetikzlibrary{mindmap}
\usetikzlibrary{decorations.pathreplacing, decorations.fractals, calligraphy}
\usetikzlibrary{decorations.pathmorphing}
\usetikzlibrary{decorations.markings}
\usetikzlibrary{decorations.text}
\usetikzlibrary{matrix}

\tikzstyle arrowstyle=[scale=1]
\tikzstyle directed=[postaction={decorate,decoration={markings,
    mark=at position .5 with {\arrow[arrowstyle]{stealth}}}}]
\tikzstyle reverse directed=[postaction={decorate,decoration={markings,
    mark=at position .5 with {\arrowreversed[arrowstyle]{stealth};}}}]
    
\definecolor{markcolor}{rgb}{.25,0,1}
\definecolor{darkpastelgreen}{rgb}{0.01, 0.75, 0.24}

\tikzstyle{vecArrow} = [thick, decoration={markings,mark=at position
   1 with {\arrow[semithick]{open triangle 60}}},
   double distance=1.4pt, shorten >= 5.5pt,
   preaction = {decorate},
   postaction = {draw,line width=1.4pt, white,shorten >= 4.5pt}]
\tikzstyle{innerWhite} = [semithick, white,line width=1.4pt, shorten >= 4.5pt]

\pgfdeclaredecoration{irregular fractal line}{init}
{
  \state{init}[width=\pgfdecoratedinputsegmentremainingdistance]
  {
    \pgfpathlineto{\pgfpoint{random*\pgfdecoratedinputsegmentremainingdistance}{(random*\pgfdecorationsegmentamplitude-0.02)*\pgfdecoratedinputsegmentremainingdistance}}
    \pgfpathlineto{\pgfpoint{\pgfdecoratedinputsegmentremainingdistance}{0pt}}
  }
}

\tikzset{
   paper/.style={draw=black!10, blur shadow, every shadow/.style={opacity=1, black}, shading=bilinear interpolation,
                 lower left=black!10, upper left=black!5, upper right=white, lower right=black!5, fill=none},
   irregular cloudy border/.style={decoration={irregular fractal line, amplitude=0.2},
           decorate,
     },
   irregular spiky border/.style={decoration={irregular fractal line, amplitude=-0.2},
           decorate,
     },
   ragged border/.style={ decoration={random steps, segment length=7mm, amplitude=2mm},
           decorate,
   }
}
    
\tikzset{
 path image/.style={
        	 path picture={
		  \node[opacity=.75] at (path picture bounding box.center) {
		    \includegraphics[width=\textwidth height=\textheight, keepaspectratio]{cmb.png}};}},
 path img/.style={
        	 path picture={
		  \node[opacity=.75] at (path picture bounding box.center) {
		    \includegraphics[width=\textwidth height=\textheight, keepaspectratio]{lss.png}};}}
}

\title{Wavefunctionals/S-matrix techniques in de Sitter}

\author*[a]{Paolo Benincasa}

\affiliation[a]{Max-Planck-Institut f{\"u}r Physik (Werner-Heisenberg-Institut),\\
  F{\"o}hringer Ring 6, 80805 M{\"u}nchen, Germany}

\emailAdd{pablowellinhouse@anche.no}

\abstract{Flat-space physics is highly constrained by basic principles such as Lorentz invariance, locality, unitarity and causality. This is neatly seen in the structure of scattering amplitudes. For processes occurring in an expanding background we do not have the same level of understanding, not even in the case of de Sitter space. In this talk I provide a brief account of some of the recent efforts which aim to fill this gap. I will discuss some novel results in the understanding of the analytic structure of the Bunch-Davies wavefunction(al) of the universe in the perturbative regime, some fundamental constraints that it ought to satisfy as well as some general insights coming from an underlying combinatorial structure.
\\
\vspace{.25cm}

\hspace{11.575cm} MPP-2022-34}

\FullConference{%
  Corfu Summer Institute 2021 "School and Workshops on Elementary Particle Physics and Gravity"\\
  29 August - 9 October 2021\\
  Corfu, Greece
}

\begin{document}
\maketitle


\section{Introduction}\label{sec:Intro}

Cosmology is our window on the physics at {\it ultra high} energies: the correlations we can observe in the Cosmic Microwave Background (CMB) and in the Large Scale Structures (LSS) can be traced back to correlations of quantum fields at end of inflation, which encode the imprint of processes occurring during the inflationary period where the Hubble scale can be as large as $10^{14}$ GeV. Understanding how to extract fundamental physics from such objects would then teach us about which processes can occur at much higher energy scales than any experiment on earth.

However, we can either write down our favourite Lagrangians in an expanding background, make our predictions and then just waiting for observation to eventually validate (or disprove) our ideas, or we can aim at understanding what are the rules governing physical processes in such a background and which constraints on the physics they impose. In this latter perspective, such rules are expected to have an imprint on the cosmological correlations in terms of their analytic structure. One might wonder why they should exist at all. First, cosmological correlations are the result of a time evolution and live at on a space-like surface at late times. If we assume that such an evolution has been causal and unitary, such principles should constrain their function form. Secondly, the cosmological correlators contain the flat-space scattering amplitudes in a particular kinematic limit \cite{Maldacena:2011nz, Raju:2012zr}, for which we do know constraints coming from unitarity and locality \cite{Weinberg:1964ew, Benincasa:2007xk, Benincasa:2011pg, McGady:2013sga, Arkani-Hamed:2017jhn}. It is then reasonable to expect that there should be an avatar of such constraints in the structure of the cosmological correlations and the latter should be constrained by the requirement of having the correct flat-space limit. Hence, given a cosmological correlation function how do we know if it really comes from a causal and unitary evolution in a cosmological space-time?

We can ask these questions at the level of the wavefunction(al) of the universe rather than on the correlators. Why should we care about the wavefunction of the universe given that it is not what we will ever measure? In a more general sense, the wavefunction can also be considered as a {\it physical observable} as it has properties such as gauge-invariance. Furthermore, it can be considered as a more primitive object as its square modulus provides the probability distributions for field configurations from which we can compute any correlation involving (gauge-invariant) operators constructed from the relevant fields:
\begin{equation}\label{eq:fav}
	\langle f\rangle\:=\:\mathcal{N} \int \mathcal{D}\,\Phi|\Psi[\Phi]|^2 f(\Phi),	
\end{equation}
where $\Psi[\Phi]$ is the wavefunction which depends on some field configuration $\Phi$ at the space-like future boundary, $f(\Phi)$ is the quantity of which we would like to compute the spatial average, and $\mathcal{N}$ is a suitable normalisation constant. Hence, the properties of $\langle f\rangle$ are (partially) inherited from the properties of $\Psi[\Phi]$. The wavefunction of the universe will be the observable with focus on, and will be chosen to be in the Bunch-Davies vacuum.

It is interesting to notice that the questions we are posing for the wavefunction have some answer for the flat-space scattering amplitudes, at least in perturbation theory. In this context, the basics rules are given by the flat-space isometry group, locality, unitarity and causality. The isometry group dictates that the scattering amplitudes should be a function of Lorentz-invariant variables and fixes, up to an overall constant, the three-particle amplitudes for any (unitary) irreducible representations, providing a (non-perturbative) expression for the smallest possible processes and a classification for the three-particle interactions \cite{Benincasa:2007xk,Arkani-Hamed:2017jhn}. Locality reflects into the analyticity of the scattering amplitudes, with at most poles and branch-cuts which are associated to the propagation of particles only. Unitarity is encoded into the factorisation properties of the scattering amplitudes as the singularities are approached, with a positivity conditions on the coefficients of such singularities. Finally, the avatar of causality is the Steinmann relations \cite{Steinmann:1960soa, Steinmann:1960sob, Araki:1961hb, Ruelle:1961rd}, {\it i.e.} the statement that the double discontinuity across partially overlapping channels has to vanish in the physical region \cite{Stapp:1971hh, Cahill:1973px, Lassalle:1974jm, Cahill:1975qp}. A simple requirement of consistency with these principles a great deal of results: the consistency of interacting theories for particles with spin less or equal to $2$ and the inconsistency for those involving a finite number of particles with higher spin \cite{Weinberg:1964ew, Benincasa:2007xk, Benincasa:2011pg, McGady:2013sga, Arkani-Hamed:2017jhn}; the charge conservation and the equivalence principle \cite{Weinberg:1964ew, Benincasa:2007xk}, the existence of non-trivial self-interactions for spin-$1$ particles just if there are different species and an internal symmetry which satisfies the Jacobi identity \cite{Benincasa:2007xk}. Interestingly, all these results do not require neither the knowledge of the putative Lagrangian nor the notion of fields: the building blocks, the three-particle amplitudes, are fixed by the isometry group and amplitudes involving a higher number of states can be constructed via consistency with these principles.

In cosmology we are pretty far from such an understanding, but in recent years several progress to fill this gap has been made. De Sitter space represents a suitable playground to start addressing these issues: its isometry group, $SO(1,d+1)$, is nothing but the conformal group in $d$-dimensions and such a symmetry fixes the three-state processes up to a constant \cite{Polyakov:1970xd, Osborn:1993cr, Maldacena:2011nz, Creminelli:2011mw, Bzowski:2013sza, Pajer:2016ieg}. We would then need to gain a deeper understanding of the analytic structure of the wavefunction and how unitarity and causality constraints it. It turns out that the Bunch-Davies wavefunction has singularities in correspondence of the vanishing of sums of energies\footnote{With an abuse of language, we refer to the modulus of a spatial momentum as {\it energy}: $E:=|\vec{p}|$.} associated to the full process as well as sub-processes: when these singularities are approached, the wavefunction reduces to the high-energy limit of the flat-space scattering amplitudes (in the former case) or factorises into a product of lower points scattering amplitudes and wavefunctions \cite{Arkani-Hamed:2017fdk, Arkani-Hamed:2018kmz}. Furthermore, on one side unitarity implies a cosmological optical theorem \cite{Goodhew:2020hob} and related cutting rules \cite{Melville:2021lst, Goodhew:2021oqg}, and on the other the wavefunction has to satisfy Steinmann-like relations and their extension to multiple singularities \cite{Benincasa:2020aoj, Benincasa:2021qcb}. While all these statements hold for the perturbative wavefunctions for general FRW cosmologies, non-perturbative implications of unitarity started to be formulated in de Sitter case \cite{Hogervorst:2021uvp, DiPietro:2021sjt}.

In this talk I will report on such recent progress, with a special focus on the perturbative Bunch-Davies wavefunction of the universe for scalar interactions as well as with a first-principle definition in terms of {\it cosmological polytopes} \cite{Arkani-Hamed:2017fdk, Benincasa:2019vqr}.


\section{The wavefunction of the universe}\label{sec:WFU}

Let us begin with the usual definition of the wavefunction of the universe. Let us consider a system described by the action $S[\phi]$, $\phi$ being the collection of modes of relevance, in an FRW background
\begin{equation}\label{eq:Bkg}
	ds^2\:=\:a^2(\eta)\,\left[-d\eta^2+d\,d\vec{x}\cdot d\vec{x}\right],
\end{equation}
which has its late-time boundary at $\eta=0$. Then the wavefunction of the universe can be written as a path integral -- see \cite{Benincasa:2022gtd}:
\begin{equation}\label{eq:WFdef}
	\Psi[\Phi]\:=\:\mathcal{N}\int\mathcal{D}\phi\,e^{i S[\phi]}\:\sim\:
		e^{-\phi\phi\psi_2-\phi\phi\phi\psi_3-\ldots-\overbrace{\phi\cdots\phi}^{\mbox{\tiny $n$-times}}\psi_n-\ldots}
\end{equation}
where $\mathcal{N}$ being a normalisation constant, and the very right-hand-side represents its general structure with $\psi_n$ being the $n$-point wavefunction coefficient. In perturbation theory it can be seen as the sum over all the Feynman graphs with a certain number of external states:
\begin{equation}\label{eq:FeynR}
	\psi_n\: :=\:\sum_{\{\mathcal{G}\}}\psi_{\mathcal{G}},\qquad\,
	\psi_{\mathcal{G}}\:=\:\delta^{\mbox{\tiny $(d)$}}\left(\sum_{j=1}^n\vec{p}^{\mbox{\tiny $(j)$}}\right)
		\int_{-\infty}^0\prod_{s\in\mathcal{V}}\left[d\eta_s\,\phi_{+}^{\mbox{\tiny $(v)$}}V_s\right]
		\prod_{e\in\mathcal{E}}G(y_e;\eta_{s_e},\eta_{s'_e}),
\end{equation}
$\psi_{\mathcal{G}}$ being the wavefunction associated to a given $n$-point graph $\mathcal{G}$ defined by the sets of sites\footnote{For avoiding later to have a language clash with the polytope terminology, we reserve the word {\it vertex} for the highest codimension boundary of a polytope, and use {\it site} for graphs.} $\mathcal{V}$ and edges $\mathcal{E}$; $\psi_+^{\mbox{\tiny $(s)$}}$ is the product of the bulk-to-boundary propagators $\phi_+$ at a site $s$, $V_s$ encodes the interaction at a site $s$; and finally $G(y_e;\eta_{s_e},\eta_{s'_e})$ is the bulk-to-bulk propagator associated to the edge $e$ and with internal energy $y_e$. The Bunch-Davies condition at in the infinite past selects those modes which are exponentially suppressed as $\eta\,\longrightarrow\,-\infty(1-i\varepsilon)$, {\it i.e.} the positive energy solutions
\begin{equation}\label{eq:BD}
	\lim_{\eta\longrightarrow-\infty(1-i\varepsilon)}\phi_+(-E\eta)\:\sim\:f(\eta)e^{iE\eta},
\end{equation}
for some function $f(\eta)$. Furthermore, the bulk-to-bulk propagator is characterised by three terms
\begin{equation}\label{eq:Gdef}
	\begin{split}
		G(y_e;\eta_{s_e},\eta_{s'_e})\:=\:\frac{1}{\mbox{Re}\{2\psi_2(y_e)\}}&
		\left[
			\overline{\phi}_+(-y_e\eta_{s_e})\phi_+(-y_e\eta_{s'_e})\vartheta(\eta_{s_e}-\eta_{s'_e})+
		\right.\\
			&\hspace{.25cm}
			+\phi_+(-y_e\eta_{s_e})\overline{\phi}_+(-y_e\eta_{s'_e})\vartheta(\eta_{s'_e}-\eta_{s_e})-\\
		&\left.\hspace{.25cm}
			-\phi_+(-y_e\eta_{s_e})\phi_+(-y_e\eta_{s'_e})
		\right]
	\end{split}
\end{equation}
where $\psi_2(y_e)$ is the two-point wavefunction with energy $y_e$. Notice that the first two terms are time-ordered, while the last one is a boundary term coming from the boundary condition that the fluctuations have to vanish at $\eta=0$. Then, given a graph $\mathcal{G}$, the associated wavefunction $\psi_{\mathcal{G}}$ has $3^{n_e}$ terms. Through \eqref{eq:FeynR} can in principle compute any perturbative contribution to the wavefunction. Beside the intrinsic difficulties of such a computation, is there anything we can say a priori without computing any of the $\psi_{\mathcal{G}}$'s and, hence, without specifying the theory?


\section{Bunch-Davies condition, singularities and factorisations}\label{sec:SingFact}

Luckily, the answer to the previous question is affirmative. As in the disclaimer at the very beginning, the focus of our discussion is the perturbative wavefunction with Bunch-Davies initial condition. As the latter selects just the positive energy solutions, the wavefunction can have singularities just as sums of certain subset of the energies vanish. There are two important consequences. First, it implies that no particle production/decay is allowed in the physical region and hence the wavefunction is analytic for combination of the energies involving differences \cite{Arkani-Hamed:2017fdk, Arkani-Hamed:2018kmz}
\begin{equation}\label{eq:AFS}
	\lim_{x-y\longrightarrow 0}\psi_{\mathcal{G}}\:=\:\mathcal{P}(x-y)
\end{equation}
where $x$ and $y$ are some linear combinations of the external and internal energies respectively, and $\mathcal{P}(x-y)$ is a polynomial in $x-y$. One says that {\it folded singularities} have to be absent.

The second important consequence of the Bunch-Davies conditions is that the allowed singularities are reachable outside the physical sheet, the latter being defined by $\{E_j,\,y_e\,\in\,\mathbb{R}_+,\;\forall\,j=1,\ldots,n,\: e\in\mathcal{E}\}$: in order to approach them, it is necessary to perform an analytic continuation such that some of the energies become negative and other stay positive. Consequently, moving outside the physical sheet in this way we have a situation with in- and out-states. Furthermore, saying that a certain sum of the energies vanishes is equivalent to say that energy conservation for a subprocess is restored: the singularities are related to the flat-space physics! More precisely:
\begin{itemize}
	\item[\ding{111}] if we consider the vanishing of the total energy $E_{\mbox{\tiny tot}}:=\sum_{j=1}^nE_j$, then the coefficient of the singularity 
		is the (high-energy limit of) the flat-space scattering amplitude with the same $n$ external states  \cite{Maldacena:2011nz, Raju:2012zr}:
		\begin{equation}\label{eq:EtotSing}
			\psi_{\mathcal{G}}\:\overset{E_{\mbox{\tiny tot}}\longrightarrow0}{\sim}\:
				\mathcal{A}_{\mathcal{G}}\,\mbox{Sing}\{E_{\mbox{\tiny tot}}\}
		\end{equation}
		A way to understand it is to consider the integral representation for $\psi_{\mathcal{G}}$ in the region where the center-of-mass time 
		$\overline{\eta}$ for all states is taken to infinite past. Because of the Bunch-Davies condition, the mode functions reduce to exponential 
		and
		\begin{equation}\label{eq:psietotlim}
			\psi_{\mathcal{G}}\:\sim\:\int_{-\infty(1-i\varepsilon)}d\overline{\eta}\,f_{\mbox{\tiny tot}}(\overline{\eta})\,
				e^{iE_{\mbox{\tiny tot}}\overline{\eta}}.
		\end{equation}
		This contribution would vanish unless $E_{\mbox{\tiny tot}}=0$ outside of the physical region, providing an energy-conserving delta function.
		Furthermore, taking the $\overline{\eta}$ to past infinity moves the interactions arbitrarily far away from the space-like boundary. For 
		states which do not have flat-space counterpart, and hence there is no notion of S-matrix, the $E_{\mbox{\tiny tot}}$ singularity is milder 
		and the coefficient is a purely cosmological effect \cite{Grall:2020ibl}.
	\item[\ding{111}] if we consider the vanishing of some {\it partial} energy 
		$E_{\mathfrak{g}}:=\sum_{j\in\mathfrak{g}}E_j+\sum_{\centernot{e}\in\centernot{\mathcal{E}}}y_{\centernot{e}}$ associated to a subgraph 
		$\mathfrak{g}\subset\mathcal{G}$ with $\centernot{\mathcal{E}}\subset\mathcal{E}$ being the subset of edges of $\mathcal{G}$ departing from 
		$\mathfrak{g}$, then the wavefunction factorises into the product of a flat-space scattering amplitude $\mathcal{A}_{\mathfrak{g}}$ associated
		to $\mathfrak{g}$ times the wavefunction associated to the complementary graph $\overline{\mathfrak{g}}$ summed over both positive and 
		negative energies associated to the edges between $\mathfrak{g}$ and $\overline{\mathfrak{g}}$ 
		\cite{Arkani-Hamed:2017fdk, Arkani-Hamed:2018kmz}:
		\begin{equation}\label{eq:EpartSing}
			\psi_{\mathcal{G}}\:\overset{E_{\mbox{\tiny $\mathfrak{g}$}}\longrightarrow0}{\sim}\:
				\mathcal{A}_{\mathfrak{g}}\times
				\sum_{\{\sigma_{\centernot{e}}=\mp\}}\frac{\psi_{\overline{\mathfrak{g}}}(\sigma_{\centernot{e}}y_{\centernot{e}})}{
					\mbox{Re}\{2\psi_2(y_{\centernot{e}})\}}\times\mbox{Sing}\{E_{\mathfrak{g}}\}.
		\end{equation}
		This factorisation property can be understood with a similar argument as \eqref{eq:psietotlim} and considering that in this limit just two out
		of the three terms of the bulk-to-bulk propagator \eqref{eq:Gdef} associated to each of the edges in the subset $\centernot{\mathcal{E}}$ 
		contribute, one of the two time-ordered terms and the boundary one.
	\item[\ding{111}] we can also consider the codimension-$2$ singularity reached by considering the energies $E_{\mathfrak{g}}$ and 
		$E_{\overline{\mathfrak{g}}}$, associated to a subgraph $\mathfrak{g}$ and its complementary $\overline{\mathfrak{g}}$, vanishing. Then energy
		conservation is restored in the two complementary subprocesses and the wavefunction factorises into a product of flat-space scattering 
		amplitudes:
		\begin{equation}\label{eq:ElErSing}
			\psi_{\mathcal{G}}\:\overset{E_{\mbox{\tiny $\mathfrak{g}$}},E_{\mbox{\tiny $\overline{\mathfrak{g}}$}}\longrightarrow0}{\sim}\:
				\mathcal{A}_{\mathfrak{g}}\times\mathcal{A}_{\overline{\mathfrak{g}}}\,\times\,
				\mbox{Sing}\{E_{\mbox{\tiny $\mathfrak{g}$}}\}\,\times\,\mbox{Sing}\{E_{\mbox{\tiny $\overline{\mathfrak{g}}$}}\}.
		\end{equation}
\end{itemize}
Notice that depending on the specific theory, as these limits are approached, the scattering amplitude can enjoy Lorentz boosts and hence be full-fledge Lorentz invariant \cite{Arkani-Hamed:2018ahb}, as in the case of states in de Sitter space,  or still be boostless and just invariant under the Euclidean group $ISO(d):=\mathbb{R}^d\rtimes SO(d)$ \cite{Pajer:2020wnj}. These factorisation properties can be used to bootstrap the four-point wavefunction in de Sitter without making any reference \cite{Arkani-Hamed:2018kmz, Baumann:2020dch} as well as to reconstruct the so-called {\it wavefunction universal integrand} for an arbitrary graph \cite{Benincasa:2018ssx}.


\section{Cosmological unitarity}\label{sec:CosmUn}

The factorisation properties of a flat-space scattering amplitude are a direct consequence of unitarity. What can we say about the imprint of unitarity in the Bunch-Davies wavefunction? The evolution operator $\hat{U}:=\mathcal{T}\{\mbox{exp}\{-i\int_{-\infty}^0d\eta\,\hat{H}(\eta)\}\}$, defined out of the Hamiltonian operator describing our system, has to satisfy the unitarity condition $\hat{U}\hat{U}^{\dagger}=\hat{\mathbb{I}}$. Leaving on a side the issue of a regularisation of $\hat{U}$ that makes it well-defined in the infinite past and, at the same time, does not spoil unitarity -- see \cite{Baumgart:2020oby, Benincasa:2022gtd} --, then requiring that the wavefunction comes from an unitary evolution implies that it has to obey an optical theorem and the related cutting rules \cite{Goodhew:2020hob, Melville:2021lst, Goodhew:2021oqg, Meltzer:2021zin}:
\begin{equation}\label{eq:ccr}
	\psi_{\mathcal{G}}+\psi_{\mathcal{G}}^{\dagger}\:=\:\sum_{\{\mathcal{E}_c\}}
	\left[
		\prod_{e\in\mathcal{E}_c}\int\frac{d^d q_{s_e}}{(2\pi)^d}\int\frac{d^d q_{s'_e}}{(2\pi)^d}\frac{1}{2\mbox{Re}\{\psi_2(y_e)\}}
	\right]
	\prod_{\mathfrak{g}\subset\mathcal{G}}\left(\psi_{\mathfrak{g}}+\psi_{\mathfrak{g}}^{\dagger}\right),
\end{equation}
which expressed the left-hand-side in terms of the sum of all the possible ways of deleting an edge, splitting the original graph $\mathcal{G}$ into a collection of subgraphs $\mathcal{G}_c$, {\it i.e.} $\mathcal{G}=\cup_{\mathfrak{g}\in\mathcal{G}_c}\mathfrak{g}$. Also, $\{\mathcal{E}_c\}$ is the collection of all the subsets of $\mathcal{E}$, excluding the empty set; $s_e$ and $s'_e$ are the endpoints of the edge $e$, and all the $\psi^{\dagger}$ have all the energies reversed in sign, except the one associated to the edge $e$. Interestingly enough, one can notice that $\psi_{\mathcal{G}}$ and $\psi_{\mathcal{G}}^{\dagger}$ on the left-hand-side of \eqref{eq:ccr} share only one singularity, the total energy pole, and their leading behaviour is the same up to a sign. This implies that when approaching the total energy singularity in the left-hand-side of \eqref{eq:ccr}, the leading Laurent coefficient get cancelled, and in fact the right-hand-side of \eqref{eq:ccr} does not have such a singularity. However, such a Laurent coefficient encodes the flat-space amplitude for the full graph $\mathcal{G}$ and \eqref{eq:ccr} does not reproduce the flat-space cutting rules. A careful use of the $i\varepsilon$-prescription produce distributional terms which precisely reproduce them \cite{Benincasa:2022cot}. Finally, prescinding on such distributional terms, \eqref{eq:ccr} together with the requirement that no folded singularities are allowed provides a new tool to bootstrap the wavefunction $\psi_{\mathcal{G}}$ \cite{Baumann:2021fxj}


\section{Steinmann-like relations}\label{sec:SteinLR}

The $i\varepsilon$-prescription, and hence the analyticity properties, are intimately related to causality. While its imprint on the wavefunction is not understood yet, in the flat-space scattering amplitude it reflects as constraints on codimension-$2$ singularities, named Steinmann relations \cite{Steinmann:1960soa, Steinmann:1960sob, Araki:1961hb, Ruelle:1961rd, Stapp:1971hh, Cahill:1973px, Lassalle:1974jm, Cahill:1975qp}. They state that the double discontinuity of the amplitudes across partially overlapping channels have to vanish. One can then ask a similar question on $\psi_{\mathcal{G}}$: are there constraints on singularities with codimension higher than one? Also in this case, the answer turns out to be affirmative \cite{Benincasa:2020aoj, Benincasa:2021qcb}. In particular, considering the singularities corresponding to two subgraphs $\mathfrak{g}_1,\,\mathfrak{g}_2\,\subset\,\mathcal{G}$ such that $\mathfrak{g}_1\cup\mathfrak{g}_2\,\neq\,\varnothing$, $\mathfrak{g}_1\cup\overline{\mathfrak{g}}_2\,\neq\,\varnothing$ and $\overline{\mathfrak{g}}_1\cup\mathfrak{g}_2\,\neq\,\varnothing$, then \cite{Benincasa:2020aoj}
\begin{equation}\label{eq:StmLkRls}
	\mbox{Disc}_{E_{\mathfrak{g}_1}}\left(\mbox{Disc}_{E_{\mathfrak{g}_2}}\psi_{\mathcal{G}}\right)\:=\:0,
\end{equation}
where the non-empty conditions on the intersections between the two subgraphs and their complementaries imply that the the channels they represent are partially overlapping. This is precisely the statement that the double discontinuities across partially overlapping channels is zero also for the wavefunction. Interestingly, in codimension-$2$ there are further conditions on double discontinuities that are absent in the flat-space case \cite{Benincasa:2021qcb}:
\begin{equation}\label{eq:DbDisc2}
	\mbox{Disc}_{E_{\mathfrak{g}_1}}\left(\mbox{Disc}_{E_{\mathfrak{g}_2}}\psi_{\mathcal{G}}\right)\:=\:0,\qquad
	\mbox{ for }
	\left\{
		\begin{array}{l}
			\mathfrak{g}_2\subset\mathfrak{g}_1\\
			\phantom{\cdots}\\
			n_{\mathfrak{g}_2}\,>\,L_{\mathfrak{g}_1}+1
		\end{array}
	\right.
	,
\end{equation}
where $n_{\mathfrak{g}_2}$ is the number of edges departing from $\mathfrak{g}_1$ and $L_{\mathfrak{g}_1}$ is the number of loops of $\mathfrak{g}_1$. There is a beautiful physical interpretation of why \eqref{eq:DbDisc2} ought to hold: the subgraph $\mathfrak{g}_1$ identifies a scattering amplitude $\mathcal{A}_{\mathfrak{g}_1}$ -- again $E_{\mathfrak{g}_1}=0$ imposes energy conservation for such a subprocess -- and taking $E_{\mathfrak{g}_2}=0$ with $\mathfrak{g}_2$ satisfying the two conditions above perform a cut on $\mathcal{A}_{\mathfrak{g}_1}$ with a non-defined energy flow! 

It is possible to extend these constraints to arbitrary codimension-$k$ singularities. Concretely, given $k$-channels identified by a collection of $k$ subgraphs $\{\mathfrak{g}_j\subset\mathcal{G},\,j=1,\ldots,k\}$, in principle there are $2^k$ intersections involving any combination of them and their complementary graphs. Considering all such intersections, except the one involving just the complementary graphs, {\it i.e.} $\bigcap_{j=1}^k\overline{\mathfrak{g}}_j$, then \cite{Benincasa:2021qcb}:
\begin{equation}\label{eq:MtDisck}
	\mbox{Disc}_{E_{\mathfrak{g}_1}}\left(\mbox{Disc}_{E_{\mathfrak{g}_2}}\left(\ldots\mbox{Disc}_{E_{\mathfrak{g}_k}}\psi_{\mathcal{G}}\right)\right)=0
\end{equation}
if and only if out of these $2^k-1$ intersections there are more then $k$ of them which are non-empty, {\it i.e.} not all of them have to be necessarily overlapping, but just any number greater than $k$ would make the multiple discontinuity vanish.

Some comments are now in order. The proof of all these Steinmann-like relations \eqref{eq:StmLkRls}, \eqref{eq:DbDisc2} and \eqref{eq:MtDisck} rely on the combinatorial formulation in terms of the cosmological polytope \cite{Arkani-Hamed:2017fdk, Benincasa:2019vqr}. As we will see in the next section, the cosmological polytopes provide a first principle definition for a wavefunction universal integrand describing certain scalar toy models -- the wavefunction universal integrand upon integration over the external energies with a suitable measure returns the wavefunction for an arbitrary FRW cosmology. This implies that strictly speaking these constraints are originally formulated as double and multiple residues of this universal integrand along channels satisfying the above conditions. However, the integration over the external energies returns polylogarithms \cite{Arkani-Hamed:2017fdk, Hillman:2019wgh} and consequently these restrictions can be promoted to restrictions on double and multiple discontinuities for the integrated wavefunction. In the case of the tree-level wavefunction, this is the end of the story. At loop level, there are the integrations over the loop momenta that need to be carried out and hence the integration over the external energies still produce an integrand. In this case, the space of functions of these integrals is not known and hence whether these constraints translate to the loop-integrated wavefunction still need to be proven and at loop level the statements on the discontinuities holds for the energy-integrated integrand. Finally, the cosmological polytope formulation describe processes with states with a flat-space counterpart: if on one side then the Steinmann-like relations not necessarily have to hold for those states which do not have a flat-space counterpart, on the other side any wavefunction which can be written in sums of scalar integrals that have a description in terms of cosmological polytopes will satisfy them.


\section{A combinatorial origin for the wavefunction}\label{sec:CWF}

In the previous sections we have reviewed novel progress in the understanding of the analytic structure of the Bunch-Davies wavefunction in de Sitter and more general FRW cosmologies. The advantage of de Sitter space-time is that, being maximally symmetric, there are more symmetries constraining the functional form of the wavefunction and that can be exploited. In this section we switch gear and consider a different, complementary, approach: we can look for an independent, well-defined, mathematical description which can turn out to have the basic properties we ascribe to the wavefunction. In other words, we can look for a mathematical object which is defined in its own right, without any reference to physics and then discover that it encodes the wavefunction. Such object do exist and go under the name of {\it cosmological polytopes} \cite{Arkani-Hamed:2017fdk, Benincasa:2019vqr}. They turn out to be in one-to-one correspondence with the graphs $\mathcal{G}$ encoding the contribution $\psi_{\mathcal{G}}$ to the wavefunction. Hence, there are two ways of introducing them: either from their intrinsic definition and then discover that it is possible to associate a Feynman graph to them; or starting from the graph and discover that there is a concrete polytope associated to them. For pedagogical reason, we will consider this second route. A second disclaimer, the cosmological polytopes encode the so-called wavefunction universal integrand: given a graph $\mathcal{G}$, we can think about the wavefunction contribution $\psi_{\mathcal{G}}$ as
\begin{equation}\label{eq:psiG2}
	\psi_{\mathcal{G}}\:=\:\int\prod_{l=1}^l\frac{d^d\ell}{(2\pi)^d}\prod_{s\in\mathcal{V}}\left[\int_{X_s}^{+\infty}dx_s\,\lambda(x_s)\right]
		\tilde{\psi}_{\mathcal{G}}(x_s,y_e(\vec{\ell}))
\end{equation}
where $\tilde{\psi}_{\mathcal{G}}(x_s,y_e(\vec{\ell}))$ is the universal integrand defined according to \eqref{eq:FeynR} by taking $\phi_{+}(-E\eta)=e^{iE\eta}$, the integration over $\vec{\ell}$ is the loop integration, $X_s$ is the sum of the energies of the external states, $\lambda(x_s)$ is a measure which encodes the specificity of the cosmology and of the valence of the interaction, {\it e.g.} in the case of a $\phi^3$ theory in $dS_{1+3}$ space we have $\lambda(x_s)=1$. For conformal theories, there is no $x_s$ integration: $\tilde{\psi}_{\mathcal{G}}$ can be also interpreted as the wavefunction in Minkowski space with a space-like boundary at $\eta=0$. The cosmological polytopes encode $\tilde{\psi}_{\mathcal{G}}(x_s,y_e(\vec{\ell}))$ and provide with a combinatorial way of extract information about the $x_s$-integrated function via the so-called symbols \cite{Arkani-Hamed:2017fdk, Hillman:2019wgh, Benincasa:2022gtd}. Finally, given a graph $\mathcal{G}$ let us suppress the external lines corresponding to the bulk-to-boundary propagators: such reduced graph $\tilde{\mathcal{G}}$ is therefore just defined by the set of sites $\mathcal{V}$ and internal edges $\mathcal{E}$ of $\mathcal{G}$. We can associate the kinematic information to $\tilde{\mathcal{G}}$ by assigning the label $x_s$ to the $s$-th site and $y_e$ to the edge $e\in\mathcal{E}$: each element of $\{x_s,\,s\in\mathcal{V}\}$ can be though of as the sum of the energies of the external states at the site $s$, while $\{y_e,\,e\in\mathcal{E}\}$ are the moduli of the momenta running in the bulk-to-bulk propagators and parametrise the loop momenta as well as the angles between all the momenta.

\begin{figure*}[t]
	\centering
 \begin{tikzpicture}[line join = round, line cap = round, ball/.style = {circle, draw, align=center, anchor=north, inner sep=0},
                     axis/.style={very thick, ->, >=stealth'}, pile/.style={thick, ->, >=stealth', shorten <=2pt, shorten>=2pt}, 
		     every node/.style={color=black},
		     scale={.75}, transform shape]
  \begin{scope}[scale={.5}, shift={(6,3)}, transform shape]
   \pgfmathsetmacro{\factor}{1/sqrt(2)};
   \coordinate  (B2) at (1.5,-3,-1.5*\factor);
   \coordinate  (A1) at (-1.5,-3,-1.5*\factor);
   \coordinate  (B1) at (1.5,-3.75,1.5*\factor);
   \coordinate  (A2) at (-1.5,-3.75,1.5*\factor);
   \coordinate  (C1) at (0.75,-.65,.75*\factor);
   \coordinate  (C2) at (0.4,-6.05,.75*\factor);
   \coordinate (Int) at (intersection of A2--B2 and B1--C1);
   \coordinate (Int2) at (intersection of A1--B1 and A2--B2);

   \tikzstyle{interrupt}=[
    postaction={
        decorate,
        decoration={markings,
                    mark= at position 0.5
                          with
                          {
                            \node[rectangle, color=white, fill=white, below=-.1 of Int] {};
                          }}}
   ]

   \draw[interrupt,thick,color=red] (B1) -- (C1);
   \draw[-,very thick,color=blue] (A1) -- (B1);
   \draw[-,very thick,color=blue] (A2) -- (B2);
   \draw[-,very thick,color=blue] (A1) -- (C1);
   \draw[-, dashed, very thick, color=red] (A2) -- (C2);
   \draw[-, dashed, thick, color=blue] (B2) -- (C2);

   \coordinate[label=below:{\Large ${\bf x'}_i$}] (x2) at ($(A1)!0.5!(B1)$);
   \draw[fill,color=blue] (x2) circle (2.5pt);
   \coordinate[label=left:{\Large ${\bf x}_i$}] (x1) at ($(C1)!0.5!(A1)$);
   \draw[fill,color=blue] (x1) circle (2.5pt);
   \coordinate[label=right:{\Large ${\bf x}_j$}] (x3) at ($(B2)!0.5!(C2)$);
   \draw[fill,color=blue] (x3) circle (2.5pt);
  \end{scope}
  \begin{scope}[scale={.5}, shift={(15,3)}, transform shape]
   \pgfmathsetmacro{\factor}{1/sqrt(2)};  
   \coordinate (B2c) at (1.5,-3,-1.5*\factor);
   \coordinate (A1c) at (-1.5,-3,-1.5*\factor);
   \coordinate (B1c) at (1.5,-3.75,1.5*\factor);
   \coordinate (A2c) at (-1.5,-3.75,1.5*\factor);  
   \coordinate (C1c) at (0.75,-.65,.75*\factor);
   \coordinate (C2c) at (0.4,-6.05,.75*\factor);
   \coordinate (Int3) at (intersection of A2c--B2c and B1c--C1c);

   \node at (A1c) (A1d) {};
   \node at (B2c) (B2d) {};
   \node at (B1c) (B1d) {};
   \node at (A2c) (A2d) {};
   \node at (C1c) (C1d) {};
   \node at (C2c) (C2d) {};

   \draw[-,dashed,fill=blue!30, opacity=.7] (A1c) -- (B2c) -- (C1c) -- cycle;
   \draw[-,thick,fill=blue!20, opacity=.7] (A1c) -- (A2c) -- (C1c) -- cycle;
   \draw[-,thick,fill=blue!20, opacity=.7] (B1c) -- (B2c) -- (C1c) -- cycle;
   \draw[-,thick,fill=blue!35, opacity=.7] (A2c) -- (B1c) -- (C1c) -- cycle;

   \draw[-,dashed,fill=red!30, opacity=.3] (A1c) -- (B2c) -- (C2c) -- cycle;
   \draw[-,dashed, thick, fill=red!50, opacity=.5] (B2c) -- (B1c) -- (C2c) -- cycle;
   \draw[-,dashed,fill=red!40, opacity=.3] (A1c) -- (A2c) -- (C2c) -- cycle;
   \draw[-,dashed, thick, fill=red!45, opacity=.5] (A2c) -- (B1c) -- (C2c) -- cycle;
  \end{scope}
  \begin{scope}[scale={.6}, shift={(5,-3.5)}, transform shape]
   \pgfmathsetmacro{\factor}{1/sqrt(2)};
   \coordinate  (c1b) at (0.75,0,-.75*\factor);
   \coordinate  (b1b) at (-.75,0,-.75*\factor);
   \coordinate  (a2b) at (0.75,-.65,.75*\factor);

   \coordinate  (c2b) at (1.5,-3,-1.5*\factor);
   \coordinate  (b2b) at (-1.5,-3,-1.5*\factor);
   \coordinate  (a1b) at (1.5,-3.75,1.5*\factor);

   \coordinate (Int1) at (intersection of b2b--c2b and b1b--a1b);
   \coordinate (Int2) at (intersection of b2b--c2b and c1b--a1b);
   \coordinate (Int3) at (intersection of b2b--a2b and b1b--a1b);
   \coordinate (Int4) at (intersection of a2b--c2b and c1b--a1b);
   \tikzstyle{interrupt}=[
    postaction={
        decorate,
        decoration={markings,
                    mark= at position 0.5
                          with
                          {
                            \node[rectangle, color=white, fill=white] at (Int1) {};
                            \node[rectangle, color=white, fill=white] at (Int2) {};
                          }}}
   ]

   \node at (c1b) (c1c) {};
   \node at (b1b) (b1c) {};
   \node at (a2b) (a2c) {};
   \node at (c2b) (c2c) {};
   \node at (b2b) (b2c) {};
   \node at (a1b) (a1c) {};

   \draw[interrupt,thick,color=red] (b2b) -- (c2b);
   \draw[-,very thick,color=red] (b1b) -- (c1b);
   \draw[-,very thick,color=blue] (b1b) -- (a1b);
   \draw[-,very thick,color=blue] (a1b) -- (c1b);
   \draw[-,very thick,color=blue] (b2b) -- (a2b);
   \draw[-,very thick,color=blue] (a2b) -- (c2b);

   \node[ball,text width=.15cm,fill,color=blue, above=-.06cm of Int3, label=left:{\large ${\bf x}_i$}] (Inta) {};
   \node[ball,text width=.15cm,fill,color=blue, above=-.06cm of Int4, label=right:{\large ${\bf x'}_i$}] (Intb) {};

  \end{scope}
  \begin{scope}[scale={.6}, shift={(12.5,-3.5)}, transform shape]
   \pgfmathsetmacro{\factor}{1/sqrt(2)};
   \coordinate (c1a) at (0.75,0,-.75*\factor);
   \coordinate (b1a) at (-.75,0,-.75*\factor);
   \coordinate (a2a) at (0.75,-.65,.75*\factor);
  
   \coordinate (c2a) at (1.5,-3,-1.5*\factor);
   \coordinate (b2a) at (-1.5,-3,-1.5*\factor);
   \coordinate (a1a) at (1.5,-3.75,1.5*\factor);

   \draw[-,dashed,fill=green!50,opacity=.6] (c1a) -- (b1a) -- (b2a) -- (c2a) -- cycle;
   \draw[draw=none,fill=red!60, opacity=.45] (c2a) -- (b2a) -- (a1a) -- cycle;
   \draw[-,fill=blue!,opacity=.3] (c1a) -- (b1a) -- (a2a) -- cycle; 
   \draw[-,fill=green!50,opacity=.4] (b1a) -- (a2a) -- (a1a) -- (b2a) -- cycle;
   \draw[-,fill=green!45!black,opacity=.2] (c1a) -- (a2a) -- (a1a) -- (c2a) -- cycle;  
  \end{scope}
  \begin{scope}[shift={(11,0)}, transform shape]
   \coordinate[label=below:{\footnotesize $x_i$}] (x1) at (0,0);
   \coordinate[label=below:{\footnotesize $x'_i$}] (x2) at ($(x1)+(1.5,0)$);
   \coordinate[label=above:{\footnotesize $y_{ii'}$}] (yi) at ($(x1)!0.5!(x2)$);
   \coordinate[label=below:{\footnotesize $x_j$}] (x3) at ($(x2)+(1.5,0)$);
   \coordinate[label=above:{\footnotesize $y_{i'j}$}] (yj) at ($(x2)!0.5!(x3)$);

   \draw[-, very thick, color=red] (x1) -- (x2) -- (x3);
   \draw[color=blue,fill=blue] (x1) circle (2pt);
   \draw[color=blue,fill=blue] (x2) circle (2pt);
   \draw[color=blue,fill=blue] (x3) circle (2pt);
  \end{scope}
  \begin{scope}[shift={(11,-3)}, transform shape]
   \coordinate[label=left:{\footnotesize $x_i$}] (x1) at (.75,0);
   \coordinate[label=right:{\footnotesize $x_j$}] (x2) at ($(x1)+(1.5,0)$);
   \coordinate (c) at ($(x1)!0.5!(x2)$);
   \coordinate[label=above:{\footnotesize $y_{ij}$}] (yi) at ($(c)+(0,.75)$);
   \coordinate[label=below:{\footnotesize $y_{ji}$}] (yj) at ($(c)-(0,.75)$);
   
   \draw[color=red, very thick] (c) circle (.75cm);
   \draw[color=blue,fill=blue] (x1) circle (2pt);
   \draw[color=blue,fill=blue] (x2) circle (2pt);
  \end{scope}
 \end{tikzpicture}
\caption{Example of cosmological polytopes (central column), their first principle definition as intersection of triangles (first column) and their associated graphs}
 \label{fig:cp}
\end{figure*}
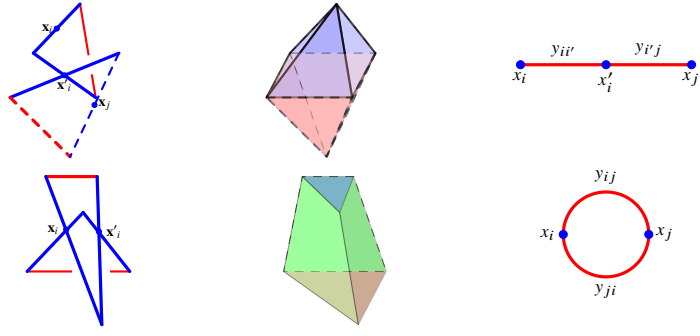
We can consider any graph $\tilde{\mathcal{G}}$ with $n_e$ edges as generated from a collection of $n_e$ graphs with two sites by suitably identifying the sites. Given a single two-site graph, we can promote the labels associated to its sites and edge, namely $x_1,\,y_{12},\,x_2$ to be local coordinates $\mathcal{Y}:=(x_1,y_{12},x_2)$ in $\mathbb{P}^2$. We can take the canonical basis of this space $\{\mathbf{x}_1,\,\mathbf{y}_{12},\,\mathbf{x}_2\}:=\{(1,0,0),\,(0,1,0),\,(0,0,1)\}$ to identify the midpoints of a triangle, {\it i.e.} there is a triangle associated to the $2$-site line graph, with vertices given by
\begin{equation}\label{eq:TrV}
	\{\mathbf{x}_1-\mathbf{y}_{12}+\mathbf{x}_2,\,\mathbf{x}_1+\mathbf{y}_{12}-\mathbf{x}_2,\,-\mathbf{x}_1+\mathbf{y}_{12}+\mathbf{x}_2\}
\end{equation}
Then, considering a collection of $n_e$ $2$-site line graphs translates in considering a collection of $n_e$ disconnected triangles embedded in $\mathbb{P}^{3n_e-1}$. Identifying two sites instead translates in intersection two sides of two triangles in their midpoints, {\it i.e.} requiring a linear combination among pairs of vertices of the two triangles, {\it e.g.} (see Figure \ref{fig:cp})
\begin{equation}\label{eq:LC}
	(-\mathbf{x}_i+\mathbf{y}_{ii'}+\mathbf{x'}_i)+(\mathbf{x}_i+\mathbf{y}_{ii'}-\mathbf{x'}_i)\:=\:\mathbf{x'}_i\:=
	(\mathbf{x'}_i-\mathbf{y}_{i'j}+\mathbf{x}_j)+(\mathbf{x'}_i+\mathbf{y}_{i'j}-\mathbf{x}_j)
\end{equation}
Hence, given an arbitrary graph $\tilde{\mathcal{G}}$ with $n_s$ sites and $n_e$ edges, we can promote all these labels to be local coordinates $\mathcal{Y}\::=\:(x_{1},\ldots,x_{n_s},y_{1},\ldots,y_{n_e})$ of the projective space $\mathbb{P}^{n_s+n_e-1}$, and the associated cosmological polytope is the convex hull of $n_e$ triples such as \eqref{eq:TrV} with sharing the suitable points $\mathbf{x}_s$.

Given a cosmological polytope, it is possible to associate a differential form, called canonical form, with logarithmic singularities only and only its boundaries
\begin{equation}\label{eq:CPCF}
	\omega(\mathcal{Y},\mathcal{P}_{\tilde{\mathcal{G}}})\:=\:
		\Omega(\mathcal{Y},\mathcal{P}_{\tilde{\mathcal{G}}})\langle\mathcal{Y}d^{n_s+n_e-1}\mathcal{Y}\rangle\:=\:
	\frac{\mathfrak{n}(\mathcal{Y})\langle\mathcal{Y}d^{n_s+n_e-1}\mathcal{Y}\rangle}{\displaystyle
					\prod_{j=1}^{\tilde{\nu}}q_j(\mathcal{Y})}
\end{equation}
where $\{q_j(\mathcal{Y}),\,j=1,\ldots,\tilde{\nu}\}$ are linear polynomials identifying the facets\footnote{The facets are the codimension-$1$ boundaries of a polytope.} of $\mathcal{P}_{\tilde{\mathcal{G}}}$, while the numerator $\mathfrak{n}(\mathcal{Y})$ is a polynomial of degree $\tilde{\nu}-n_s-n_e$ which identify the locus of the intersections, outside of $\mathcal{P}_{\tilde{\mathcal{G}}}$, of the hyperplanes containing the facets. It has the property that the residue with respect to any of the $q_j$'s is still a canonical form which is associated to the facet identified by $q_j=0$.

It turns out that the rational function $\Omega(\mathcal{Y},\mathcal{P}_{\tilde{\mathcal{G}}})$ in \eqref{eq:CPCF}, called {\it canonical function}, is the wavefunction universal integrand $\tilde{\psi}_{\mathcal{G}}$ \cite{Arkani-Hamed:2017fdk}. This implies that the cosmological polytope $\mathcal{P}_{\tilde{\mathcal{G}}}$ codifies all the information about such an integrand as well as a great deal of information about the integrated wavefunction: the final function depends on two basic data, {\it i.e.} the integrand and the contour of integration. The integrals over the labels of $x_s$ turns out to provide a representation for the Aomoto polylogarithms \cite{Arkani-Hamed:2017fdk, Arkani-Hamed:2017ahv, Hillman:2019wgh, Benincasa:2022gtd}. Going back to the universal integrand $\tilde{\psi}_{\mathcal{G}}$, its identification with the canonical function implies that any of its properties can be formulated in terms of the combinatoric properties $\mathcal{P}_{\tilde{\mathcal{G}}}$. For example, the Steinmann-like relations discussed in the previous section emerge as compatibility conditions on the facets, {\it i.e.} conditions that the facets have to satisfy in order for them to intersect and form a higher-codimension face of $\mathcal{P}_{\tilde{\mathcal{G}}}$. If these conditions are not satisfied, than the hyperplanes containing them have an intersection outside the polytope in that codimension. Consequently, such an intersection identifies a zero of $\Omega(\mathcal{Y},\mathcal{P}_{\tilde{\mathcal{G}}})$, {\it i.e.} it is a condition to identify the numerator $\mathfrak{n}(\mathcal{Y})$.

There is a special facet which corresponds to taking the residue with respect to the total energy: the canonical form of this facet turns out to encode the flat-space scattering amplitudes. It is referred to as {\it scattering facet}. The analysis of the vertex structure of this facet allowed to provide a combinatorial prove of the flat-space cutting rules \cite{Arkani-Hamed:2018ahb} and of the actual Steinmann relations \cite{Benincasa:2020aoj}, providing a precise setting in which flat-space unitarity and flat-space causality {\it emerge} from the cosmological context. The same occurs with for Lorentz invariance \cite{Arkani-Hamed:2018ahb}.
 
It is also remarkable that the Bunch-Davies condition is encoded into the combinatorial automorphism group, {\it i.e.} the symmetry group preserving the face lattice of $\mathcal{P}_{\tilde{\mathcal{G}}}$ -- the face lattice is a lattice whose vertices are {\it all} the faces of $\mathcal{P}_{\tilde{\mathcal{G}}}$, including the full $\mathcal{P}_{\tilde{\mathcal{G}}}$ and the empty set, and the edges of the lattice are determined by containment relations. Even more remarkably, the knowledge of the scattering facet and of this combinatorial automorphisms allow to reconstruct the full polytope \cite{Benincasa:2018ssx}. Said differently, knowing the flat-space amplitude and the Bunch-Davies condition it is possible to bootstrap the wavefunction universal integrand. 


\section{Conclusion}

Quantum field theory in de Sitter space, and more generally in expanding universes, has still a lot of mysteries that need to be unravelled. An approach which does not rely on an explicit time evolution has the power of providing new insights in the physics encoded in the relevant observables. In this sense, the progress in the S-matrix context can be a source of inspiration, given that it was possible to relate many physical properties to just few fundamental assumptions. Hence, ideally it is desirable to bring our understanding of cosmological observables on the same footing as the flat-space scattering amplitudes. Understanding the analytic structure of the wavefunction of the universe and how fundamental physics is encoded into it, therefore seems to be a promising starting point: the wavefunction of the universe is a primitive object from which we can compute any other observable, and appears to have a simpler structure.

In this talk I have reviewed the first progress in addressing the following questions: what's the imprint of basic principles, such as unitarity and causality, in the wavefunction of the universe? what are their consequences?

This program is still at the very beginning and there are very basic and important issues to be addresses. All the discussion in this talk is based on a graph-by-graph analysis, while the actual observables are given by sums of graphs. While some properties carry over the sum over graphs, {\it e.g.} the cosmological optical theorem and the Steinmann-like relations, the individual graph viewpoint is a big limitation in understanding structures for general theories, especially theories which in the usual field-theoretical description would have gauge redundancies. It is desirable to have an approach, an organisation of perturbation theory, which is gauge-invariant at all steps. This has been one of the keys in several achievements in scattering amplitudes. Also the combinatorial picture is still primitive, despite the interesting results achieved: it also suffer of the same problem of being tied to a single graph.

Finally, these S-matrix-inspired approaches can be helpful to elucidate the long standing issue of the IR divergences and eventual inconsistency of perturbation theory in de Sitter -- see for example\footnote{For some brief discussions and a more complete set of references see \cite{Baumann:2022jpr, Benincasa:2022gtd}.} \cite{Ford:1984hs, Antoniadis:1985pj, Tsamis:1994ca, Tsamis:1996qm, Tsamis:1997za, Polyakov:2007mm, Polyakov:2009nq, Gorbenko:2019rza, Mirbabayi:2019qtx}.


\acknowledgments

I would like to warmly thank Dionysios Anninos, Tarek Anous, Frederik Denef, Guilherme Pimentel and Rachel Rosen for the stimulating environment created at the {\it Workshop on Quantum Features in a de Sitter Universe}. I would also like to thank Dionysios Anninos, Carlos Duaso Pueyo, Victor Gorbenko, Austin Joyce, Diego Hofman, Ruben Monten, Guilherme Pimentel for stimulating discussions. Finally, I would like to thank the developers of {\tt Polymake} \cite{polymake:2000, polymake:2017}, {\tt TOPCOM} \cite{Rambau:TOPCOM-ICMS:2002}, {\tt SageMath} \cite{sagemath}, {\tt Maxima} \cite{maxima} and {\tt Tikz} \cite{tantau:2013a}. 

This research received funding from the European Research Council (ERC) under the European Union’s Horizon 2020 research and innovation programme (grant agreement No 725110), {\it Novel structures in scattering amplitudes}.


\bibliographystyle{utphys}
\bibliography{cprefs}



\end{document}